\documentclass[twocolumn,secnumarabic,amssymb, nobibnotes, aps, prl, showpacs, superscriptaddress]{revtex4-2}

\usepackage{graphicx}
\usepackage{dcolumn}
\usepackage{bm}
\usepackage{hyperref}
\usepackage{subfiles}

\usepackage{color} %\textcolor{declared-color}{text}
\begin{document}

%\preprint{}

\title{Fermi surface and nested magnetic breakdown in WTe$_2$}

\author{J.~F.~Linnartz}
\affiliation{High Field Magnet Laboratory (HFML-EMFL), Radboud University, Toernooiveld 7, Nijmegen 6525 ED, Netherlands.}
\affiliation{Radboud University, Institute for Molecules and Materials, Nijmegen 6525 AJ, Netherlands.}

\author{C.~S.~A.~M\"uller}
\affiliation{High Field Magnet Laboratory (HFML-EMFL), Radboud University, Toernooiveld 7, Nijmegen 6525 ED, Netherlands.}
\affiliation{Radboud University, Institute for Molecules and Materials, Nijmegen 6525 AJ, Netherlands.}

\author{Yu-Te~Hsu}
\affiliation{High Field Magnet Laboratory (HFML-EMFL), Radboud University, Toernooiveld 7, Nijmegen 6525 ED, Netherlands.}
\affiliation{Radboud University, Institute for Molecules and Materials, Nijmegen 6525 AJ, Netherlands.}

\author{C.~Breth~Nielsen}
\affiliation{Department of Chemistry and iNANO, Aarhus University, Langelandsgade 140, DK-8000 Aarhus C, Denmark}

\author{M. Bremholm}
\affiliation{Department of Chemistry and iNANO, Aarhus University, Langelandsgade 140, DK-8000 Aarhus C, Denmark}

\author{N.~E.~Hussey}
\affiliation{High Field Magnet Laboratory (HFML-EMFL), Radboud University, Toernooiveld 7, Nijmegen 6525 ED, Netherlands.}
\affiliation{Radboud University, Institute for Molecules and Materials, Nijmegen 6525 AJ, Netherlands.}
\affiliation{H.H. Wills Physics Laboratory, University of Bristol, Tyndall Avenue, BS8 1TL, United Kingdom}

\author{A.~Carrington}
\email{a.carrington@bristol.ac.uk}
\affiliation{H.H. Wills Physics Laboratory, University of Bristol, Tyndall Avenue, BS8 1TL, United Kingdom}

\author{S.~Wiedmann}
\email{steffen.wiedmann@ru.nl}
\affiliation{High Field Magnet Laboratory (HFML-EMFL), Radboud University, Toernooiveld 7, Nijmegen 6525 ED, Netherlands.}
\affiliation{Radboud University, Institute for Molecules and Materials, Nijmegen 6525 AJ, Netherlands.}

\date{\today}
\begin{abstract}
We report a detailed Shubnikov-de Haas (SdH) study on the Weyl type-II semimetal WTe$_2$ in magnetic fields up to 29~T. By using the SdH results to guide our density functional theory calculations, we are able to accurately determine its Fermi surface by employing a moderate Hubbard $U$ term, which is an essential step in explaining the unusual electronic properties of this much studied material. In addition to the fundamental orbits, we observe magnetic breakdown, which can consistently be explained within the model of a `Russian-doll-nested' Fermi surface of electron and hole pockets. The onset of magnetic breakdown in WTe$_2$ is solely determined by impurity damping in contrast to magnetic breakdown scenarios in other metallic systems.
\end{abstract}


\begin{titlepage}
\maketitle
\end{titlepage}

%\section{Introduction}
The layered transition-metal dichalcogenide WTe$_2$ provides a versatile platform to investigate topologically non-trivial phases
\cite{Pan2018}. Bulk WTe$_2$ in the $T_d$ phase has been identified as the prototypical example of a Weyl type-II semimetal
\cite{Soluyanov2015,Armitage2018}. In the past few years, particular attention has been paid to the observation of a large, quadratic
and non-saturating magnetoresistance (MR) up to 60~T, a signature of a perfectly compensated semimetal \cite{Ali2014}. Bound by
van der Waals interaction, two-dimensional layers of WTe$_2$ can also be exfoliated down to a monolayer. Through conventional gating
techniques, various phases of matter such as the quantum spin Hall effect \cite{Wu2018}, superconductivity in a monolayer
\cite{Sajadi2018} and room-temperature vertical ferroelectricity in its bi- and trilayer form \cite{Fei2018} have been observed. 

The Fermi surface (FS) of bulk WTe$_2$ has been investigated through angle-resolved photoemission spectroscopy (ARPES)
\cite{Pletikosic2014,Wu2015,Jiang2015,Wu2016,Wang2016,DiSante2017,Wu2017,Das2019} and quantum oscillation (QO) experiments
\cite{Wu2017,Wu2015,Zhu2015,Cai2015,Rhodes2015,Pan2015,Rana2018,Onishi2018,Jo2019}. With previous density functional theory
(DFT) band-structure calculations, however, it has proved challenging to correctly explain the size and shape of the small Fermi
pockets \cite{Ali2014,Pletikosic2014,Jiang2015,Wu2015,Wang2016,DiSante2017,Das2019,Zhu2015,Cai2015,Rhodes2015,Pan2015,Jo2019}.

The material has two mirror symmetries and consequently, there are two groups of electron and holes pockets, see Fig.~\ref{FIG2}(d). The
majority of experimental investigations conclude that the FS does indeed comprise two pairs of electronlike pockets
and two pairs of holelike pockets, the latter located near the center of the Brillouin zone, sandwiched between the two pairs of electron
pockets. Due to the lack of inversion symmetry in WTe$_2$, the spin-orbit interaction leads to a Dresselhaus spin splitting of the bands
and hence the electron and hole FS consist of nested `Russian-doll' pairs \cite{Zhu2015,Jo2019}. A question remains, however, regarding
the exact location of the Fermi level and whether the hole pockets are connected over the $\Gamma$-point forming a `dog-bone-shaped' hole
orbit \cite{Pan2015,Jiang2015} or if additional electron pockets reside in the vicinity of the $\Gamma$-point \cite{Wu2017}.
 
One way to solve this long-standing puzzle is to exploit the phenomenon of magnetic breakdown (MB), i.e. tunneling of quasi-particles
between distinct (adjacent) pockets of the FS above a threshold magnetic field. This leads to the formation of new orbits consisting of
linear combinations of the fundamental frequencies of the associated pockets and thereby constraining further the topology of the FS.
Indeed, previous QO studies have shown that MB might occur in WTe$_2$ \cite{Zhu2015,Cai2015,Rhodes2015,Onishi2018}, though a robust
identification of these orbits has not yet been done. Additionally, O'Brien $et~al.$ proposed that magnetic breakdown in WTe$_2$ occurs
between one electron and one hole pocket due to the specific cone structure of a Weyl type-II semimetal whereby electron and hole bands
touch each other in $k$-space \cite{OBrian2016}. Knowing whether this scenario is actually manifested in WTe$_2$ depends on a detailed
knowledge of the Fermi surface, which is currently lacking.
 
In this Letter, we present high-field magneto-transport experiments on a mm-sized WTe$_2$ sample down to $^3$He temperatures. Our
analysis of the observed Shubnikov-de Haas (SdH) oscillations, when combined with our DFT calculations determines the precise electronic
structure of WTe$_2$ close to the Fermi level. Our subsequent analysis confirms the nested Russian-doll arrangement but suggests that
MB can only occur between pockets of the same sign. Moreover, and in contrast with other metallic systems, the onset of MB is found to
be solely determined by impurity damping.

\begin{figure}[ht]
	\includegraphics[width=8.5cm]{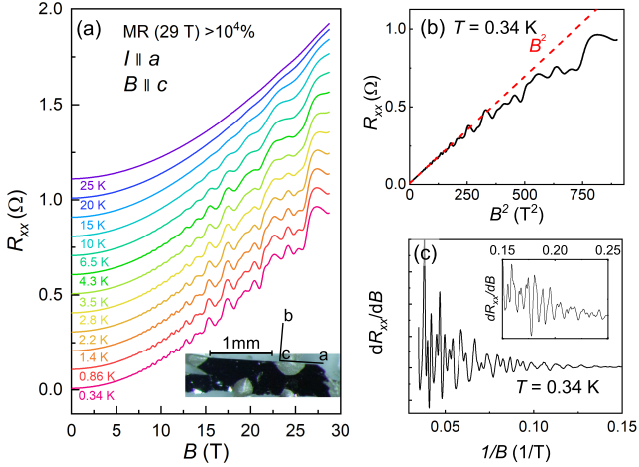}
	\caption{\label{fig1} (Color online) (a) Resistance $R_{xx}$ as a function of magnetic field measured for various temperatures up to 25~K. The curves are offset vertically for clarity. The inset shows the mm-sized exfoliated sample. (b) $R_{xx}$ plotted versus $B^2$ to highlight the quadratic behavior of the MR. (c) d$R_{xx}$/d$B$ as a function of $1/B$ at 0.34~K. The inset shows the low-field part of d$R_{xx}$/d$B$($1/B$).}
	\label{FIG1}
\end{figure}

\begin{figure}[ht]
\includegraphics[width=8cm]{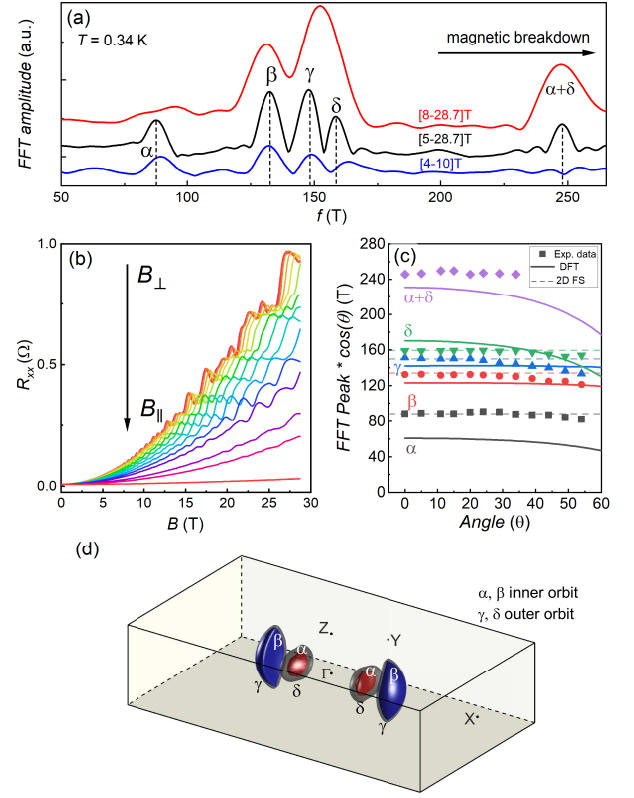}
\caption{\label{FIG2} (Color online) (a) FFT spectrum (Hann window) of the SdH signal for various magnetic field ranges. At low $B$, the individual pockets of the FS are observed. The orbits are labelled with Greek letters. With increasing field range, MB occurs and an additional orbit with frequency $f_{\alpha+\delta}$ emerges. (b) $R_{xx}$ for different tilt angles with 5$^{\circ}$ increments at $T$ = 1.4~K. (c) Rotation from (001) to (010). Maxima of the FFT peaks as a function of tilt angle are indicated with solid symbols. The dashed lines correspond to the 1/cos($\theta$) dependence expected for a two-dimensional system, the solid lines to the DFT calculation. (d) Three-dimensional FS of WTe$_2$ obtained via LDA+U DFT calculation with $U$ = 3~eV.}
\end{figure}

WTe$_{2}$ flakes were grown by vapor transport method using iodine as the transport agent. Tungsten (99.9\%) and tellurium (99.9999\%) powders were ground together, pressed to a pellet and pre-reacted in an evacuated silica tube at 750 $^{\circ}$C for two days. The product was then reground, pressed into a pellet and loaded into a silica tube with iodine pieces. The tube was placed off-center in a tube furnace to create a temperature gradient of 50 $^{\circ}$C between 850 and 800 $^{\circ}$C and left to grow for two weeks.
Figure~\ref{FIG1}(a) shows the longitudinal resistance $R_{xx}$ for the sample under study (residual resistance ratio RRR = 30) for various temperatures with the magnetic field $B$ applied along the $c$-axis while the current is passed along the $a$-axis of the sample. For all temperatures, QOs are superimposed on a $B^2$ MR of $\sim$12000 \% at maximum field. The quadratic MR persists up to 15~T indicating perfect charge carrier compensation before crossing over to $B^{1.7}$ dependence up to maximum field, see Fig.~\ref{FIG1}(b). In Fig.~\ref{FIG1}(c), d$R_{xx}$/d$B$ is plotted as a function of 1/$B$ in order to highlight the emergence of QOs for $B$~\textgreater ~4~T at $T$~=~0.34~K.

In Fig.~\ref{FIG2}(a), fast Fourier Transforms (FFTs) of the d$R_{xx}$/d$B$ data of Fig.~\ref{FIG1}(c), are presented for different magnetic field ranges. For the lowest field range [4-10]T, four distinct peaks are observed which we mark as $\alpha$, $\beta$, $\gamma$ and $\delta$ following the labelling of Ref.~\cite{Zhu2015}. For details, we refer to Fig. S1 in the Supplemental Material \cite{SI}. Using the Onsager relation, we relate the frequencies $f$ observed in the FFT spectrum to the extremal areas $A_f$ of the individual pockets: $f~=~(\hbar/2\pi e) A_f$ \cite{Shoenberg1984}. Above a threshold magnetic field ($B \geq$~8~T), the QO pattern becomes more complex. In addition to the well-developed low-frequency peaks, a high frequency orbit at 240~T appears (see the [5-28.7]T spectrum in Fig.~\ref{FIG2}(a)). This frequency corresponds approximately to the sum of the individual frequencies of the $\alpha$ and $\delta$ orbits, $f_{\alpha+\delta} = f_{\alpha} + f_{\delta}$ and is thus a viable signature of MB that will be discussed later.

The angle dependence of the SdH oscillations is presented in Fig.~\ref{FIG2}(b). $B_{\perp}$ can be determined by maximizing $R_{xx}$ due to the sensitivity of the MR of WTe$_2$ to changes in tilt angle \cite{Thoutam2015}. A commercial Hall probe has been used to extract precise values for the other tilt angles. Both the MR and the QO amplitudes strongly decrease as the sample is rotated from $B_{\perp}$ (i.e. $B$ parallel to the $c$ axis) to $B_{\parallel}$ ($B$ parallel to the $a$ axis). In order to map the FS of WTe$_2$, the frequencies of peaks obtained from the FFT spectrum [5-28.7]T multiplied by cos $\theta$ are plotted as a function of tilt angle $\theta$ in Fig.~\ref{FIG2}(c) to better visualize the quasi-2D nature of the pockets. With increasing $\theta$, the QO amplitudes of the individual orbits become more and more damped and eventually vanish above 55$^{\circ}$. The MB orbit, $f_{\alpha+\delta}$ can be observed up to $\theta$~=~30$^{\circ}$, as reported previously \cite{Zhu2015}. The persistence of this orbit over such a wide angular range contrasts markedly with recent observations of MB in nodal-line semimetals \cite{Pezzini2018,vanDelft2018,Mueller2020} where the MB orbits are found to vanish exponentially with increasing tilt angles.

To determine the Fermi surface from our extremal orbit size and mass data we use GGA+U DFT calculations \cite{Blaha2011,Perdew1996} with an experimentally determined structure \cite{Mar1992}. The calculated Fermi surface is shown in Fig.~\ref{FIG2}(d) and Fig.~\ref{FIG4}(e) and consists of two pairs of hole pockets and two pairs of electron pockets which are symmetry distinct. Each pair of electron and hole pockets is nested within each other and the surfaces are symmetrically reflected in the $\Gamma$-Z-Y plane. From this Fermi surface we identify the $\alpha$, and $\delta$ orbits as originating from the two hole pockets and $\beta$ and $\delta$ from the two electron pockets. Each FS pocket is found to have a single extremal orbit. For such small FS volumes, small shifts in the band-energies are often needed to get perfect agreement with experiment. However, for WTe$_2$, we find that in addition a moderate $U \simeq 3$\,eV must also be employed for the tungsten d orbitals in order to explain the observed value of $m_c$ for the hole pockets. Without this, the calculated mass is a factor $\sim$2 too high, even after shifting the band energies to match the observed frequencies. Physically, $U$ models the on-site Coulomb repulsion; an approach which has also been employed for example for the related material MoTe$_2$ where a similar Hubbard $U$ value was found \cite{Aryal2019,Hu2020}. Although the agreement with experiment is not perfect, the residual differences [see Table \ref{TAB1} and Fig.~\ref{FIG2}(c)] are  acceptably small. Further details of the calculations, including how the results depend on $U$, are shown in the Supplemental Material \cite{SI}.

\begin{figure}[ht]
\includegraphics[width=8cm]{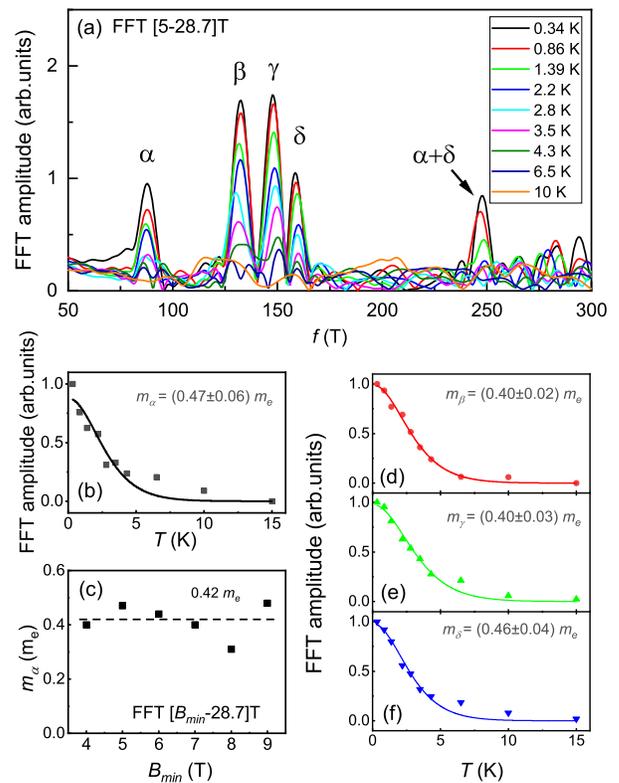}
\caption{\label{FIG3} (Color online) (a) FFT spectra in the field range [5-28.7]T for different temperatures up to 10~K for $B~||~c$. (b) Extraction of $m_c$ for the $\alpha$-pocket and (c) its corresponding range analysis in different field ranges by varying $B_{min}$. (d)-(f) Extraction of $m_c$ for the other individual pockets $\beta$, $\gamma$ and $\delta$ of the FS.}
\end{figure}

Extraction of the cyclotron masses, $m_c$, of the individual pockets of the FS is proven to be a challenge for the $\beta$, $\gamma$ and $\delta$ peaks as they are separated by only $\simeq$10~T in the FFT spectrum, and a proper range has to be chosen to avoid convolution of their FFT peaks. In Fig.~\ref{FIG3}(a), the range [5-28.7]T of the FFT spectrum, taken from d$R_{xx}$/d$B$, is plotted for different temperatures up to 10~K. Figs.~\ref{FIG3}(b), \ref{FIG3}(d), \ref{FIG3}(e), \ref{FIG3}(f) display the FFT amplitudes of the individual peaks as a function of temperature fitted to the temperature-dependent term of the Lifshitz-Kosevich formula, $R_T = X/$sinh[$X]$ with $X = \frac{2\pi ^2 k_{B}}{\hbar e}\frac{m^*_{c}T}{<B>}$, where $<1/B>=\frac{(1/B_{min})+(1/B_{max})}{2}$. Additional details concerning the mass extraction are discussed in the Supplemental Material \cite{SI}. The experimentally extracted cyclotron masses $m_c$ and those obtained from DFT calculations for each individual pocket are summarized in Table \ref{TAB1}. In order to separate the FFT peaks a relatively large field range [5-28.7]\,T is needed, which could in principle lead to an underestimate of the masses. However, the variation in $m_c$ for the $\alpha$ pocket on the field range shown in Fig.~\ref{FIG3}(c), along with additional analysis presented in the Supplemental Material \cite{SI} shows this effect is within our errors here.

\begin{table}[t]
	\centering
	\caption{Identified orbits, extracted frequencies and cyclotron masses from the experiment in the range from [5-28.7]T compared to the frequencies and cyclotron masses from the DFT calculations.}
	\label{TAB1}
	\begin{tabular} {|c|c|c|c|c|c|}
		\hline
		Orbit		& Type		& $f$ (T) 			& $m_c~(m_e)$  			& $f_{DFT}$ (T)	& $m_{c;DFT}~(m_e)$  	\\
		\hline
		\hline
		$\alpha$ 	&h			& \textbf{88}   	& $0.47\pm 0.06$ 		&	66 			& 0.43 			 		\\ 
		\hline
		$\beta$ 	&e			& \textbf{133}   	& $0.40\pm 0.02$ 		&	134 		& 0.38					\\
		\hline
		$\gamma$ 	&e			& \textbf{148}  	& $0.40\pm 0.03$		&	155 		& 0.35					\\
		\hline
		$\delta$ 	&h    		& \textbf{158}  	& $0.46\pm 0.04$		&	186 		& 0.58 				 	\\
		\hline	
	\end{tabular}
\end{table}

A further constraint of the FS topology can be made by analysing the MB orbits. Magnetic breakdown occurs if the applied magnetic field $B > B_0 \approx (h/e) (k_g/2)^2$, where $k_g$ is the breakdown gap in $k$-space. From our DFT calculations, we extract a gap of $B_0$ = 0.2~T for electrons and 0.5~T for holes, respectively. This implies that the Dingle field is around one order of magnitude larger than $B_0$, see Supplemental Material \cite{SI}. Therefore, MB is determined by impurity damping in WTe$_2$. In Fig.~\ref{FIG4}(a), the FFT spectra obtained from d$R_{xx}$/d$B$ containing all of the observed MR orbits are shown for several temperatures between 0.34 and 6.5~K in the field range [8-28.7]T. In addition to the $\alpha$+$\delta$ orbit, we identify four more MB orbits: $\beta$+$\gamma$, 2$\alpha$+$\delta$, $\alpha$+2$\delta$, 3$\alpha$+$\delta$. Given our experimental resolution, we are only able to extract the cyclotron masses of the $\alpha + \delta$, $\beta+\gamma$ and $2\alpha+\delta$ orbits, each of which corresponds to the sum of the masses of the individual pockets \cite{Shoenberg1984,Kaganov1983,Pezzini2018,vanDelft2018,Mueller2020}. Their absolute values are summarized in Table II and the corresponding fits are presented in Figs.~\ref{FIG4}(b)-(d) in which the $R_T$-term has been fitted to the $T$-dependent FFT amplitudes. 

\begin{figure}[ht]
	\includegraphics[width=8cm]{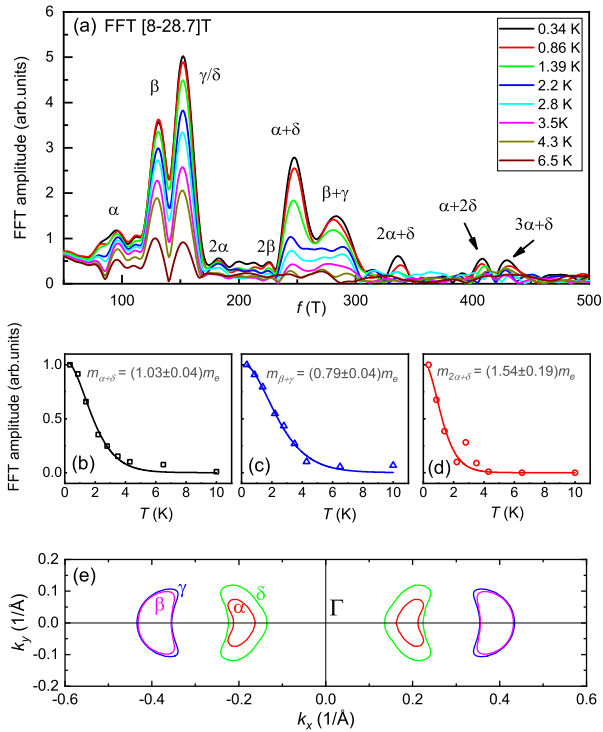} 
	\caption{\label{FIG4} (Color online) 
		(a) FFT spectrum for the field range [8-28.7]T for different temperatures up to 6.5~K. The corresponding individual and breakdown orbits are labelled. (b, c, d) FFT amplitude as a function of $T$ to extract $m_c$ of the MB orbits with the frequencies $f_{\alpha+\delta}$, $f_{\beta+\gamma}$ and $f_{2\alpha+\delta}$. (e) Cut of the FS in the $k_x-k_y$ plane highlighting the nested Russian doll configuration.}
\end{figure}

The appropriateness of the MB scenario in WTe$_2$ including the extracted cyclotron masses of the breakdown orbits and the persistence of the $\alpha+\delta$ orbit to large tilt angles, is corroborated by our DFT calculations. Knowing the topology of the Fermi surface and having determined the masses of the breakdown orbits enables us now to elucidate the origin of MB in WTe$_2$ and thus the precise FS. The scenarios based on electron-hole MB \cite{OBrian2016} and the possibility of hole-hole tunneling across the $\Gamma$ point \cite{Zhu2015} can be excluded due to the observed breakdown frequencies and (large) finite gaps in momentum space, see Fig.~\ref{FIG4}(e), making quasi-particle tunneling within our magnetic field range impossible. Another key observation is the absence of the 2$\delta$ pocket in our FFT spectra. If tunneling occured over the $\Gamma$ point, QOs with a frequency $f_{2\delta}$ would be easily observed in the FFT spectra.

In light of these considerations, the most likely scenario that is consistent with all our experimental findings is the one in which magnetic breakdown occurs between Russian doll nested electron or hole pockets in WTe$_2$. This scenario, supported by our DFT calculations allows for either electron-electron ($\beta$+$\gamma$) or hole-hole ($\alpha$+$\delta$) tunneling, see Fig. S7 \cite{SI}, within the nested pockets giving rise to frequencies of the MB orbits that we observe in the FFT spectra. The $\alpha$ pocket is situated inside the $\delta$-pocket with a similar curvature along one direction $k$ space (see Fig.~\ref{FIG4}(e)) enabling quasiparticles undergo magnetic breakdown. This scenario explains why this orbit survives out to large tilt angles as the tunneling gap will not markedly change with increasing $\theta$.

\begin{table}[t]
	\centering
	\caption{Magnetic breakdown orbits and their extracted cyclotron masses using the field fitting range [8-28.7]T.}
	\label{tab2}
	\begin{tabular} {|c|c|c|c|}
		\hline
		Orbit			& $f$ (T) 			& $m_c~(m_e)$  			& 	type				 	\\
		\hline
		\hline
		$\alpha+\delta$ & \textbf{247}   	& $1.03\pm 0.04$ 		&	hole-hole				\\ 
		\hline
		$\beta+\gamma$ 	& \textbf{283}    	& $0.79\pm 0.04$ 		&	electron-electron		\\
		\hline
		$2\alpha+\delta$ & \textbf{336}   	& $1.54\pm 0.19$		&	hole-hole				\\
		\hline	
	\end{tabular}
\end{table}

In conclusion, we have performed magneto-transport experiments up to 29~T combined with DFT calculations that enable us to precisely determine the Fermi surface of WTe$_2$. Four individual pockets and their corresponding cyclotron masses have been identified and extracted. The Fermi surface exhibits quasi-2D behavior upon tilting the magnetic field away from the $c$ axis of the crystal. All observed orbits originating from magnetic breakdown have been assigned that further constrains the topology of the Fermi surface. The unprecedented resolution of our high-field study enables us to extract the cyclotron masses of MB orbits and together with the longevity of one of the MB orbits at finite tilt angles, allows for an unambiguous determination of magnetic breakdown in WTe$_2$ that is seen to occur between two electron and hole pockets in a nested Russian-doll configuration. 

\begin{acknowledgments}
This work was supported by HFML-RU/NWO-I, a member of the European Magnetic Field Laboratory (EMFL) and by the UK Engineering and Physical Sciences Research Council (Grant No. EP/R011141/1). This publication is part of the project TOPCORE (OCENW.GROOT.2019.048) of the research program NWO - GROOT which is financed by the Dutch Research Council (NWO). We gratefully acknowledge funding from the VILLUM FOUNDATION via the Centre of Excellence for Dirac Materials (11744). M.B. acknowledges the Danish Council for Independent Research, Natural Sciences under the Sapere Aude program (Grant No.702700077B) We thank Malte R\"{o}sner and Kamran Behnia for helpful discussions.
\end{acknowledgments}

\clearpage
%\title{SI of Fermi surface and nested magnetic breakdown in WTe$_2$}
%\maketitle
%\setcounter{figure}{0}
%\setcounter{table}{0}  
%\renewcommand{\thetable}{S\arabic{table}}  
%\renewcommand{\thefigure}{S\arabic{figure}}
%\subfile{WTe2_SI_Subfile.tex}
\end{document}